\documentclass{article}
\usepackage{spconf,amsmath,graphicx}
\usepackage{booktabs}
\usepackage{multirow}
\usepackage[table,xcdraw]{xcolor}
\usepackage{hyperref}
\newcommand{\squeezeup}{\vspace{-1.6mm}}

\title{Transformer-Based Multi-Aspect Multi-Granularity \\ Non-native English Speaker Pronunciation Assessment}
\name{Yuan Gong$^1$, Ziyi Chen$^2$, Iek-Heng Chu$^2$, Peng Chang$^2$, James Glass$^1$}
\address{$^1$MIT CSAIL, Cambridge, MA 02139, USA \quad\quad $^2$PAII Inc., Palo Alto, CA 94306, USA \\
\begin{small} \texttt{\{yuangong,glass\}@mit.edu} \quad\quad \texttt{\{chenziyi253,zhuyixing276,changpeng805\}@pingan.com.cn} \end{small}
\squeezeup\squeezeup}

\begin{document}
%
\maketitle
\begin{abstract}
Automatic pronunciation assessment is an important technology to help self-directed language learners. While pronunciation quality has multiple aspects including accuracy, fluency, completeness, and prosody, previous efforts typically only model one aspect (e.g., accuracy) at one granularity (e.g., at the phoneme-level). In this work, we explore modeling multi-aspect pronunciation assessment at multiple granularities. Specifically, we train a Goodness Of Pronunciation feature-based Transformer (GOPT) with multi-task learning. Experiments show that GOPT achieves the best results on speechocean762 with a public automatic speech recognition (ASR) acoustic model trained on Librispeech. Code at \href{https://github.com/YuanGongND/gopt}{\color{blue}{https://github.com/YuanGongND/gopt}}.
\end{abstract}

\squeezeup\squeezeup
\section{Introduction}
\squeezeup
\label{sec:intro}
Computer assisted pronunciation training (CAPT) is an important technology for self-directed language learning~\cite{Eskenazi2009review, Zechner2009SpeechRater,witt2012automatic}, which facilitates non-native (L2) speakers to learn foreign spoken (L1) languages. Compared with conventional classes, CAPT is more economical and convenient, and also allows language learners to receive immediate feedback on their pronunciation. Due to its usefulness, CAPT has been extensively studied, with the majority of these efforts focusing on scoring phoneme-level pronunciation quality (e.g.,~\cite{Witt2000GOP,zhang2008automatic,luo2009analysis,wang2012improved,hu2015improved,Shi2020ContextawareGOP,van2010using}).  Overall pronunciation quality includes many other aspects such as word- and utterance-level fluency, prosody, stress, etc., which have been typically modeled separately (e.g.,~\cite{cucchiarini1998quantitative,cucchiarini2000quantitative,bagshaw1994automatic,tepperman2005automatic,arias2010automatic, LI2017innotation}). However, phoneme-, word-, and utterance-level scores of accuracy, fluency, prosody, and stress are potentially correlated, therefore modeling them jointly instead of separately may allow a machine learning model to learn a more comprehensive representation and in turn improve its performance. In reality, it is also desirable to have a \emph{single} model that can assess \emph{multiple} aspects of pronunciation simultaneously.

As a step in this direction, in this paper we propose a new pronunciation assessment model, named GOPT, based on Goodness of Pronunciation (GOP) features and a Transformer self-attention architecture~\cite{vaswani2017attention}. We use the open-source speechocean762 dataset~\cite{zhang2021speechocean762} that contains one phoneme-level, three word-level, and five utterance-level labels including accuracy, prosody, and fluency and apply \emph{multi-aspect multi-grained} supervision for GOPT training. This not only enables GOPT to measure multiple aspects of pronunciation quality, but also boosts its performance for each assessment task. In addition, the Transformer architecture captures the contextual information between phonemes and words of an utterance. As a consequence, GOPT noticeably outperforms previous methods on the speechocean762 benchmark for both phoneme- and utterance-level assessment tasks (there is no previous work reporting word-level scores). To our knowledge, this is the first work studying multi-aspect L2 speaker pronunciation assessment in a multi-granularity fashion. 

\begin{figure*}[t]
  \includegraphics[width=17.5cm]{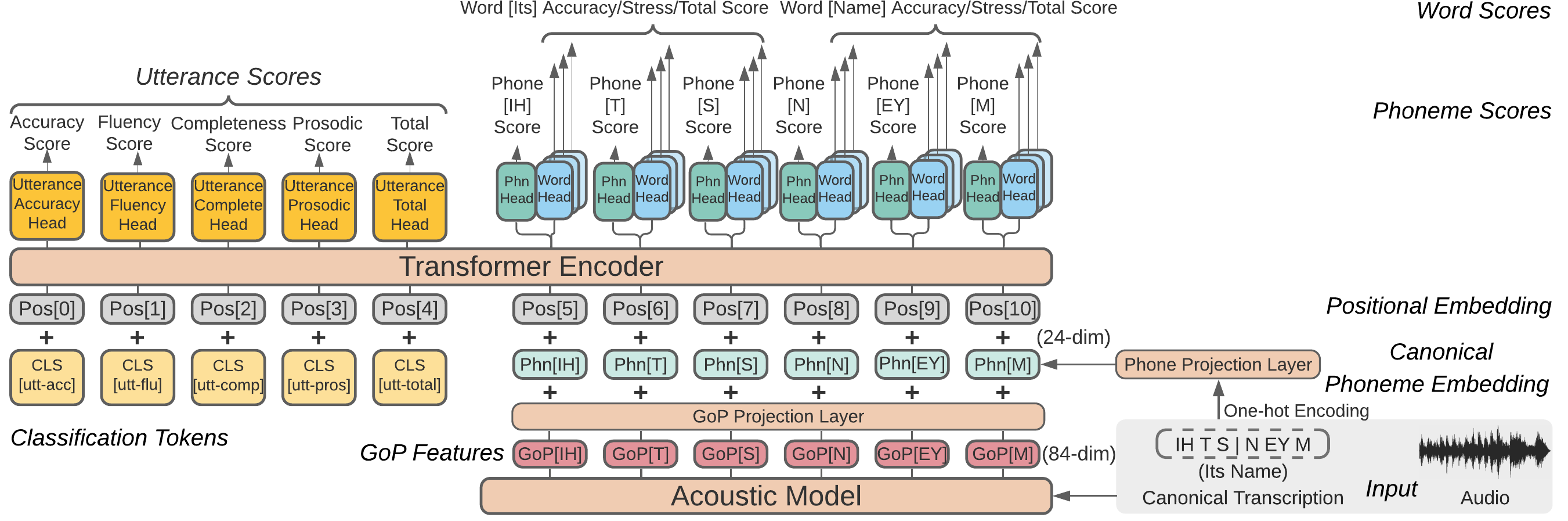}
  \caption{Illustration of the proposed GOPT architecture with a sample utterance ``Its Name'', actual utterances used are longer.}
  \label{fig:arc}
  \squeezeup\squeezeup
\end{figure*}

\squeezeup\squeezeup
\section{Related Work}
\squeezeup

As mentioned, CAPT has been extensively studied with a long history. One major focus of this area is automatic mispronunciation detection, where GOP~\cite{Witt2000GOP} and its variants (e.g.,~\cite{van2010using,zhang2008automatic,wang2012improved,hu2015improved}) are dominant methods. To capture the correlation between phonemes and words of an utterance, self-attention based models such as Transformer~\cite{vaswani2017attention} have been added on top of GOP features for score modeling to improve performance~\cite{Shi2020ContextawareGOP,lin2020automatic}. There are also some non-GOP based methods such as a wav2vec2-based method~\cite{xu2021explore} and a deep feature based method~\cite{lin2021deep} where transfer learning is usually needed due to the limited L2 training material.

Conversely, automatic assessment of other aspects of pronunciation quality are usually modeled independently, e.g., fluency~\cite{cucchiarini1998quantitative,cucchiarini2000quantitative}, prosody~\cite{bagshaw1994automatic,tepperman2005automatic}, intonation~\cite{arias2010automatic, LI2017innotation}. There are only a few previous efforts on multi-granularity pronunciation assessment~\cite{lin2020automatic,cincarek2009automatic}. In these works, however, only a single score is considered for each granularity. In addition, the hierarchical architecture in~\cite{lin2020automatic} requires a relatively sophisticated training scheme to optimize. 

To the best of our knowledge, this paper is the first to simultaneously consider multiple pronunciation quality aspects (accuracy, fluency, prosody, etc) along with multiple granularities (phoneme, word, utterance). In addition, we show that a BERT-style~\cite{devlin2018bert} non-hierarchical standard Transformer architecture can perform well on most assessment tasks. Unlike many previous efforts using non-public datasets or acoustic models, in this work, we intentionally use a public acoustic model and dataset for our main experiments (which achieves state-of-the-art results) for easy reproduction and comparison.

\squeezeup\squeezeup
\section{Goodness of Pronunciation Transformer}
\squeezeup
\label{sec:method}

\subsection{Speechocean762 Dataset}
Speechocean762~\cite{zhang2021speechocean762} is a free open-source dataset designed for pronunciation assessment, consisting of a total of 5,000 English utterances collected from 250 non-native speakers. One major advantage of speechocean762 is that it provides rich label information. Specifically, for each utterance, it provides five utterance-level aspect scores: accuracy, fluency, completeness, prosody, and total score (ranging from 0-10). For each word, it provides three word-level aspect scores: accuracy, stress, and total score (ranging from 0-10). It also provides an accuracy score for each phoneme (ranging from 0-2). Each score is annotated by five experts. Thus, it provides a total of 8 labels for different granularities and pronunciation quality aspects. However, the rich annotation has not been fully utilized by previous work. We re-scale utterance and word-level scores to 0-2, making them on the same scale as the phoneme scores. The training set consists of 2,500 utterances, 15,849 words, and 47076 phones; the test set consists of 2,500 utterances, 15,967 words, and 47,369 phones.

\begin{table*}[]
\small
\centering
\setlength\tabcolsep{4.3pt}
\begin{tabular}{lcccccccccc}
\hline
\multicolumn{1}{c}{}                        & \multicolumn{2}{c}{Phoneme  Score}                                                                                                                             & \multicolumn{3}{c}{Word Score (PCC)}                                                                                                                                                                                                          & \multicolumn{5}{c}{Utterance Score (PCC)}                                                                                                                                                                                                                                                                                                                                                                     \\ \cline{2-11} 
\multicolumn{1}{c}{\multirow{-2}{*}{Model}} & \multicolumn{1}{l}{MSE $\downarrow$}                                                       & \multicolumn{1}{c|}{PCC $\uparrow$}                                                     & Accuracy $\uparrow$                                                                      & Stress $\uparrow$                                                                     & \multicolumn{1}{c|}{Total $\uparrow$}                                                    & Accuracy $\uparrow$                                                                      & Completeness $\uparrow$                                                                  & Fluency $\uparrow$                                                                       & Prosodic $\uparrow$                                                                      & Total $\uparrow$                                                                         \\ \hline
RF~\cite{zhang2021speechocean762}                               & 0.130                                                                         & 0.440                                                                        & -                                                                             & -                                                                             & -                                                                             & -                                                                             & -                                                                             & -                                                                             & -                                                                             & -                                                                             \\
SVR~\cite{zhang2021speechocean762}                              & 0.160                                                                         & 0.450                                                                        & -                                                                             & -                                                                             &                                                                               & -                                                                             & -                                                                             & -                                                                             & -                                                                             & -                                                                             \\
Lin et.al~\cite{lin2021deep}                                   & -                                                                             & -                                                                            & -                                                                             & -                                                                             & -                                                                             & -                                                                             & -                                                                             & -                                                                             & -                                                                             & 0.720                                                                         \\ \hline
LSTM                                        & \begin{tabular}[c]{@{}c@{}}0.089\\ $\pm$0.000\end{tabular}                        & \begin{tabular}[c]{@{}c@{}}0.591\\ $\pm$0.003\end{tabular}                       & \begin{tabular}[c]{@{}c@{}}0.514\\ $\pm$0.003\end{tabular}                        & \begin{tabular}[c]{@{}c@{}}\textbf{0.294}\\ $\pm$\textbf{0.012}\end{tabular}                        & \begin{tabular}[c]{@{}c@{}}0.531\\ $\pm$0.004\end{tabular}                        & \begin{tabular}[c]{@{}c@{}}\textbf{0.720}\\ $\pm$\textbf{0.002}\end{tabular}                         & \begin{tabular}[c]{@{}c@{}}0.076\\ $\pm$0.086\end{tabular}                        & \begin{tabular}[c]{@{}c@{}}0.745\\ $\pm$0.002\end{tabular}                        & \begin{tabular}[c]{@{}c@{}}0.747\\ $\pm$0.005\end{tabular}                        & \begin{tabular}[c]{@{}c@{}}0.741\\ $\pm$0.002\end{tabular}                        \\ \cline{2-11} 
\begin{tabular}[c]{@{}l@{}}GOPT \\ (Librispeech)\end{tabular}                                        & \textbf{\begin{tabular}[c]{@{}c@{}}0.085\\ $\pm$0.001\end{tabular}}               & \textbf{\begin{tabular}[c]{@{}c@{}}0.612\\ $\pm$0.003\end{tabular}}              & \textbf{\begin{tabular}[c]{@{}c@{}}0.533\\ $\pm$0.004\end{tabular}}               & \begin{tabular}[c]{@{}c@{}}0.291\\ $\pm$0.030\end{tabular}                        & \textbf{\begin{tabular}[c]{@{}c@{}}0.549\\ $\pm$0.002\end{tabular}}               & \begin{tabular}[c]{@{}c@{}}0.714\\ $\pm$0.004\end{tabular}                        & \textbf{\begin{tabular}[c]{@{}c@{}}0.155\\ $\pm$0.039\end{tabular}}               & \textbf{\begin{tabular}[c]{@{}c@{}}0.753\\ $\pm$0.008\end{tabular}}               & \textbf{\begin{tabular}[c]{@{}c@{}}0.760\\ $\pm$0.006\end{tabular}}               & \textbf{\begin{tabular}[c]{@{}c@{}}0.742\\ $\pm$0.005\end{tabular}}               \\ \cline{2-11} 
{\color[HTML]{656565} \begin{tabular}[c]{@{}l@{}}GOPT \\ (PAII-A)\end{tabular}}       & {\color[HTML]{656565} \begin{tabular}[c]{@{}c@{}}0.069\\ $\pm$0.000\end{tabular}} & {\color[HTML]{656565} \begin{tabular}[c]{@{}c@{}}0.679\\ $\pm$0.001\end{tabular}} & {\color[HTML]{656565} \begin{tabular}[c]{@{}c@{}}0.588\\ $\pm$0.004\end{tabular}} & {\color[HTML]{656565} \begin{tabular}[c]{@{}c@{}}0.146\\ $\pm$0.004\end{tabular}} & {\color[HTML]{656565} \begin{tabular}[c]{@{}c@{}}0.601\\ $\pm$0.003\end{tabular}} & {\color[HTML]{656565} \begin{tabular}[c]{@{}c@{}}0.727\\ $\pm$0.004\end{tabular}} & {\color[HTML]{656565} \begin{tabular}[c]{@{}c@{}}0.011\\ $\pm$0.069\end{tabular}} & {\color[HTML]{656565} \begin{tabular}[c]{@{}c@{}}0.692\\ $\pm$0.015\end{tabular}} & {\color[HTML]{656565} \begin{tabular}[c]{@{}c@{}}0.694\\ $\pm$0.009\end{tabular}} & {\color[HTML]{656565} \begin{tabular}[c]{@{}c@{}}0.732\\ $\pm$0.006\end{tabular}} \\ \hline
\end{tabular}
\caption{Comparing the performance of various pronunciation assessment tasks between GOPT and baseline models. GOPT (PAII-A) depends on a different acoustic model so its results (shown in grey) cannot be directly compared with other models.}
\label{tab:mainres}
\squeezeup
\end{table*}

\squeezeup\squeezeup
\subsection{GOPT Architecture Overview}
An overview of the GOPT architecture is shown in Figure~\ref{fig:arc}. For the pronunciation assessment task, the canonical transcription is known. We first input the audio and corresponding canonical transcription to the acoustic module to get a sequence of frame-level phonetic posterior-probabilities, which are then force-aligned at the phoneme-level and converted to 84-dimensional goodness of pronunciation (GOP) features (discussed in Section~\ref{sec:am}). The GOP feature is then projected to 24-dimensions with a dense layer. In parallel, we generate a sequence of canonical phoneme embeddings (also at the phoneme-level) by first converting each canonical phoneme to a one-hot encoding and then projecting it to the same 24-dimensions as the projected GOP feature. The reason for using a canonical phoneme embedding is because different phonemes have different characteristics and thus the canonical phoneme provides useful information to the Transformer model~\cite{lin2021deep}. We then add the projected GOP feature, canonical phoneme embedding, and a 24-dimensional trainable positional embedding together and input it to the Transformer encoder. For simplicity, we intentionally follow the original Transformer encoder architecture~\cite{vaswani2017attention} as close as possible but scale it down to 3 layers and an embedding dimension of 24. 

Unlike previous work~\cite{lin2020automatic,lin2021deep} that use a hierarchical architecture to get utterance level representations, we prepend a set of five trainable \texttt{[cls]} tokens to the phoneme-level input sequence in a similar way as BERT~\cite{devlin2018bert}, each corresponding to one utterance aspect label, and use the output of the Transformer encoder of these \texttt{[cls]} aspect tokens as the corresponding utterance-level representations. The reason why this regime works is that the Transformer can learn the correlation between the utterance-level tokens and phoneme-level tokens through the attention mechanism.

During training we apply multi-task learning to the model. Specifically, we use one regression head for each phoneme, word, and utterance label (eight in total). Each regression head is a $24\times1$ dense layer with layer normalization. Utterance-level regression heads are added on top of the output of the Transformer of the corresponding utterance \texttt{[cls]} tokens. Phoneme- and word-level regression heads are added on top of the Transformer output of each corresponding phoneme. We propagate the word score to each of its phonemes during training and average the output of phonemes that belong to the word in inference. We use mean squared error (MSE) loss for each assessment task. Since we normalize the scores to the same scale, for simplicity, we first average the losses of each granularity and then sum them up with the same weight, i.e., $\mathcal{L} = \mathcal{L}_{utterance} + \mathcal{L}_{word}+ \mathcal{L}_{phoneme}$, where $\mathcal{L}_{utterance}$ and $\mathcal{L}_{word}$ are averaged utterance and word level losses of five utterance-level labels and three word-level labels, respectively; $\mathcal{L}_{phoneme}$ is the phoneme loss. The entire network (except the acoustic model) is trained end-to-end.

\subsection{Acoustic Model and GOP Feature}
\label{sec:am}

For our main experiment we use a public ASR acoustic model\footnote{https://kaldi-asr.org/models/m13} trained with Librispeech~\cite{panayotov2015librispeech} 960-hour data. The model is based on the factorized time-delay neural network (TDNN-F) and trained with the Kaldi Librispeech S5 recipe. 

Acoustic model trained on both L1 and L2 speech generates better alignment for L2 speech and may output better GOP features~\cite{tu2018investigating}. To explore if GOPT works with different acoustic models, we also test with two PAII internal acoustic models PAII-A and PAII-B, both are also TDNN-F models. PAII-A is trained with 452 hours L1 TED-LIUM 3~\cite{hernandez2018ted} data and 1,696 hours of L2 data collected from 5,994 non-native speakers; PAII-B is trained with 995 hours of L1 data (from WSJ~\cite{marcus1993building}, TED-LIUM 3~\cite{hernandez2018ted}, and Librispeech~\cite{panayotov2015librispeech}) and 6,591 hours of L2 data from 672k non-native speakers.


In this work, we use the log phone posterior (LPP) and log posterior ratio (LPR) defined in~\cite{hu2015improved} as GOP features. Specifically, the LPP of a phone $p$ is defined as follows:
\begin{equation}
    LPP(p) \approx \frac{1}{t_e-t_s+1} \sum_{t=t_s}^{t_e}\log p(p|o_t)
\end{equation}
\begin{equation}
    p(p|o_t) = \sum_{s \in p} p(s|o_t)
\end{equation}
where $t_s$ and $t_e$ are the start and end frame indexes; $o_t$ is the input observation of the frame $t$, $s$ is the state belonging to the phone $p$. LPR of a phone $p_j$ versus $p_i$ is defined as:
\begin{equation}
    LPR(p_j|p_i) = \log p(p_j|\mathbf o; t_s, t_e) - \log p(p_i|\mathbf o; t_s, t_e)
\end{equation}
The Librispeech acoustic model we use has a total of 42 pure phones, thus the GOP feature of phone $p$ can be defined as a 84-dimensional vector as follows:
\begin{equation}
    {[LPP(p_1)...,LPP(p_{42}), LPR(p_1|p)..., LPR(p_{42}|p)]}
\end{equation}

\squeezeup\squeezeup
\section{Experiments}
\squeezeup

For all experiments, we train the model with an Adam optimizer, an initial learning rate of 1e-3, a batch size of 25, and MSE loss for 100 epochs using the official speechocean762 training set, and evaluate on the official test set. The learning rate is cut in half every five epochs after the 20th epoch, and the result of the last epoch is reported. We repeat each experiment five times with different random seeds and report the mean and standard deviation of the results. Since the speechocean762 labels are imbalanced (biased towards high scores), we use the Pearson correlation coefficient (PCC) as the main evaluation metric but also report MSE of the phoneme accuracy score to make a comparison with previous work. Note that while we re-scale the utterance and word level scores, PCCs and phoneme-level MSE are not impacted. 

\subsection{Main Results}

We compare the following six models: 1) Random forest regression (RF) model implemented in the code repository of~\cite{zhang2021speechocean762}; 2) Support vector regressor (SVR) based model in~\cite{zhang2021speechocean762}; 3) Deep feature and transfer learning-based model presented in~\cite{lin2021deep}; 4) An LSTM based model implemented by us. To make a fair comparison, the LSTM model has the same depth and embedding dimension as the GOPT model and is trained with the same setting. The output of the last token is used as the utterance representation and, as with GOPT, the word score is propagated to its phones; 5) The proposed GOPT model with the Librispeech acoustic model. 6) The proposed GOPT model with the PAII-A acoustic model. It is worth mentioning that models 1-5 are all based on acoustic models trained with the same Librispeech data, and models 1,2,4,5, and 6 use the same GOP features (model 3 does not use GOP features but deep transfer learning). Therefore, we make a fair comparison and the performance difference is not due to the acoustic model and GOP features.

We show the results in Table~\ref{tab:mainres}. The key findings are as follows: First, the proposed GOPT model can perform well on most assessment tasks except word stress score and sentence completeness score assessment, demonstrating that it is possible to have a single model for multi-aspect and multi-granularity pronunciation assessment. Specifically, the GOPT achieves 0.085 MSE and 0.612 PCC for the phoneme accuracy score assessment, noticeably outperforming the models in~\cite{zhang2021speechocean762}; GOPT achieves 0.742 PCC for the utterance-level score assessment, noticeably outperforming the model in~\cite{lin2021deep} which uses more sophisticated features than GOP. We hypothesize that the poor utterance completeness assessment performance is due to the highly imbalanced distribution of the completeness score in the training data. Second, the multi-task learning scheme can be also applied to an LSTM, which achieves similar results for utterance assessment with GOPT.  However, the performance of the LSTM for phoneme-level and word-level assessment are worse than the GOPT, demonstrating that the Transformer architecture is better at modeling fine-grained pronunciation units. Third, using the PAII-A acoustic model trained on both L1 and L2 speech can further boost the phoneme and word assessment performance by around 10\%, but the utterance-level performance is worse than just using the Librispeech acoustic model. We also evaluate GOPT with PAII-B acoustic model, it leads to similar results with GOPT with PAII-A acoustic model.

\begin{table}[t]
\centering
\small
\setlength\tabcolsep{4.3pt}
\begin{tabular}{@{}lccc@{}}
\toprule
Setting         & Phoneme                & Word                 & Utterance            \\ \midrule
\multicolumn{4}{l}{\textit{Training Task}}                                                \\ \midrule
Only Phoneme      & 0.605$\pm$0.002          & -                     & -                     \\
Only Word       & -                     & 0.536$\pm$0.004          & -                     \\
Only Utterance  & -                     & -                     & 0.736$\pm$0.011          \\
Joint*          & \textbf{0.612$\pm$0.003} & \textbf{0.549$\pm$0.002} & \textbf{0.742$\pm$0.005} \\ \midrule
\multicolumn{4}{l}{\textit{Canonical Phoneme Embedding}}                                            \\ \midrule
w/o Phn Embed   & 0.512=0.006           & 0.472$\pm$0.006          & 0.719=0.002           \\
w/ Phn Embed*    & \textbf{0.612$\pm$0.003} & \textbf{0.549$\pm$0.002} & \textbf{0.742$\pm$0.005} \\ \midrule
\multicolumn{4}{l}{\textit{\# Transformer Layer (ASR params not included in $\#$params)}}                                       \\ \midrule
3* (27K Params)  & \textbf{0.612$\pm$0.003} & \textbf{0.549$\pm$0.002} & \textbf{0.742$\pm$0.005} \\
6 (48K Params)  & 0.605$\pm$0.003          & 0.543$\pm$0.004          & 0.731$\pm$0.003          \\ \midrule
\multicolumn{4}{l}{\textit{Embedding Dimension (ASR params not included in $\#$params)}}                            \\ \midrule
12 (8K Params)  & 0.608$\pm$0.003          & 0.544$\pm$0.008          & 0.741$\pm$0.011          \\
24* (27K Params) & \textbf{0.612$\pm$0.003} & \textbf{0.549$\pm$0.002} & \textbf{0.742$\pm$0.005} \\
48 (94K Params) & 0.605$\pm$0.003          & 0.545$\pm$0.006          & 0.738$\pm$0.004          \\
96 (355K Params) & 0.586$\pm$0.006          & 0.530$\pm$0.006          & 0.725$\pm$0.004          \\ \bottomrule
\end{tabular}
\caption{The ablation results, we only show the PCC of phoneme, word, and utterance total scores due to space limitation. * denotes the setting used in the base GOPT model.}
\label{tab:ablation}
\end{table}

\begin{table}[!t]
\small
\setlength\tabcolsep{4.0pt}
\begin{tabular}{@{}ccccccc@{}}
\toprule
\multirow{3}{*}{\begin{tabular}[c]{@{}c@{}}Scoring \\ Model\end{tabular}} & \multicolumn{6}{c}{Acoustic Model}                                                                                                                                                                                                                                                                                                                  \\ \cmidrule(l){2-7} 
                                                                          & \multicolumn{2}{c}{Librispeech}                                                                                 & \multicolumn{2}{c}{PAII-A}                                                                                      & \multicolumn{2}{c}{PAII-B}                                                                                      \\ \cmidrule(l){2-7} 
                                                                          & MSE $\downarrow$                                                   & PCC $\uparrow$                                                    & MSE $\downarrow$                                                   & PCC $\uparrow$                                                   & MSE $\downarrow$                                                   & PCC $\uparrow$                                                   \\ \cmidrule(l){2-7} 
\multicolumn{1}{l}{SVR}                                              & 0.160                                                   & 0.450                                                  & 0.118                                                   & 0.538                                                   & 0.115                                                   & 0.561                                                   \\ \cmidrule(l){2-7} 
\multicolumn{1}{l}{GOPT}                                                  & \begin{tabular}[c]{@{}c@{}}0.085\\ $\pm$0.001\end{tabular} & \begin{tabular}[c]{@{}c@{}}0.612\\ $\pm$0.003\end{tabular} & \begin{tabular}[c]{@{}c@{}}0.069\\ $\pm$0.000\end{tabular} & \begin{tabular}[c]{@{}c@{}}0.679\\ $\pm$0.001\end{tabular} & \begin{tabular}[c]{@{}c@{}}0.071\\ $\pm$0.001\end{tabular} & \begin{tabular}[c]{@{}c@{}}0.662\\ $\pm$0.001\end{tabular} \\ \bottomrule
\end{tabular}
\caption{Comparing the phoneme assessment performance between the SVR based~\cite{zhang2021speechocean762} model and proposed GOPT model with various acoustic models.}
\label{tab:expam}
\end{table}

\subsection{Ablations}

We conduct a set of ablation studies to show the performance impact of various factors. We set the GOPT model mentioned in Section~\ref{sec:method} with three Transformer layers, embedding dimension of 24, canonical phoneme embedding, and trained with all phoneme, word, and utterance assessment tasks as the base GOPT model, and then change one factor at a time to observe the performance change.

We show the results in Table~\ref{tab:ablation}. First, we see that the GOPT trained with multi-task learning achieves better results than any single-task learning model, demonstrating that multi-task learning not only allows the model to conduct multi-aspect and multi-granularity pronunciation assessment simultaneously, but also improves the performance of each individual task. Second, we see that the canonical phoneme embedding is crucial to the performance as the model trained without it performs much worse for all tasks. However, it is worth mentioning that canonical phoneme embedding is not the reason why GOPT outperforms previous methods since canonical phoneme embedding is also used in~\cite{lin2021deep}. In~\cite{zhang2021speechocean762}, each phoneme has a separate classifier, which serves a similar function as a canonical phoneme embedding. Third, we explore the performance impact of the size of GOPT model, and see that increasing either the width or depth of the network cannot further improve the performance, indicating that a small model is preferred with the relatively small dataset. Further, although the GOP feature is 84-dimensional, we show that an embedding size of 24 is sufficient to represent pronunciation quality with a Transformer.

Finally, in Table~\ref{tab:expam}, we compare the phoneme assessment performance between the SVR~\cite{zhang2021speechocean762} model and the proposed GOPT model with various acoustic models. We show that the proposed GOPT consistently leads to a significant performance improvement regardless of the acoustic model, demonstrating that the GOPT is model agnostic and can be used with different acoustic models.

\section{Conclusion}

In this paper, we present the Transformer-based multi-aspect multi-granularity pronunciation assessment model GOPT. We show that with the multi-task learning scheme, a single GOPT model can conduct multiple pronunciation tasks simultaneously, and its performance is better than the same model trained with a single task. Experiments show the GOPT can noticeably outperform previous methods on speechocean762.


\newpage
\bibliographystyle{IEEEbib}
\bibliography{main}

\begin{thebibliography}{10}

\bibitem{Eskenazi2009review}
Maxine Eskenazi,
\newblock ``An overview of spoken language technology for education,''
\newblock {\em Speech Communication}, 2009.

\bibitem{Zechner2009SpeechRater}
Klaus Zechner, Derrick Higgins, et~al.,
\newblock ``Automatic scoring of non-native spontaneous speech in tests of
  spoken english,''
\newblock {\em Speech Communication}, 2009.

\bibitem{witt2012automatic}
Silke~M Witt,
\newblock ``Automatic error detection in pronunciation training: Where we are
  and where we need to go,''
\newblock in {\em International Symposium on Automatic Detection on Errors in
  Pronunciation Training}, 2012.

\bibitem{Witt2000GOP}
S.M Witt and S.J Young,
\newblock ``Phone-level pronunciation scoring and assessment for interactive
  language learning,''
\newblock {\em Speech Communication}, 2000.

\bibitem{zhang2008automatic}
Feng Zhang, Chao Huang, et~al.,
\newblock ``Automatic mispronunciation detection for mandarin,''
\newblock in {\em ICASSP}, 2008.

\bibitem{luo2009analysis}
Dean Luo, Yu~Qiao, et~al.,
\newblock ``Analysis and utilization of mllr speaker adaptation technique for
  learners' pronunciation evaluation,''
\newblock in {\em Interspeech}, 2009.

\bibitem{wang2012improved}
Yow-Bang Wang and Lin-Shan Lee,
\newblock ``Improved approaches of modeling and detecting error patterns with
  empirical analysis for computer-aided pronunciation training,''
\newblock in {\em ICASSP}, 2012.

\bibitem{hu2015improved}
Wenping Hu, Yao Qian, et~al.,
\newblock ``Improved mispronunciation detection with deep neural network
  trained acoustic models and transfer learning based logistic regression
  classifiers,''
\newblock {\em Speech Communication}, 2015.

\bibitem{Shi2020ContextawareGOP}
Jiatong Shi, Nan Huo, and Qin Jin,
\newblock ``Context-aware goodness of pronunciation for computer-assisted
  pronunciation training,''
\newblock in {\em Interspeech}, 2020.

\bibitem{van2010using}
Joost van Doremalen, C~Cucchiarini, and H~Strik,
\newblock ``Using non-native error patterns to improve pronunciation
  verification,''
\newblock in {\em Interspeech}, 2010.

\bibitem{cucchiarini1998quantitative}
Catia Cucchiarini, Helmer Strik, and LWJ Boves,
\newblock ``Quantitative assessment of second language learners' fluency: an
  automatic approach,''
\newblock in {\em ICSLP}, 1998.

\bibitem{cucchiarini2000quantitative}
Catia Cucchiarini, Helmer Strik, and Lou Boves,
\newblock ``Quantitative assessment of second language learners’ fluency by
  means of automatic speech recognition technology,''
\newblock {\em The Journal of the Acoustical Society of America}, 2000.

\bibitem{bagshaw1994automatic}
Paul~Christopher Bagshaw,
\newblock {\em Automatic prosodic analysis for computer aided pronunciation
  teaching},
\newblock Ph.D. thesis, 1994.

\bibitem{tepperman2005automatic}
Joseph Tepperman and Shrikanth Narayanan,
\newblock ``Automatic syllable stress detection using prosodic features for
  pronunciation evaluation of language learners,''
\newblock in {\em ICASSP}, 2005.

\bibitem{arias2010automatic}
Juan~Pablo Arias, Nestor~Becerra Yoma, and Hiram Vivanco,
\newblock ``Automatic intonation assessment for computer aided language
  learning,''
\newblock {\em Speech Communication}, 2010.

\bibitem{LI2017innotation}
Kun Li, Xixin Wu, and Helen Meng,
\newblock ``Intonation classification for l2 english speech using
  multi-distribution deep neural networks,''
\newblock {\em Computer Speech and Language}, 2017.

\bibitem{vaswani2017attention}
Ashish Vaswani, Noam Shazeer, et~al.,
\newblock ``Attention is all you need,''
\newblock in {\em Advances in Neural Information Processing Systems}, 2017.

\bibitem{zhang2021speechocean762}
Junbo Zhang, Zhiwen Zhang, et~al.,
\newblock ``speechocean762: An open-source non-native english speech corpus for
  pronunciation assessment,''
\newblock in {\em Interspeech}, 2021.

\bibitem{lin2020automatic}
Binghuai Lin, Liyuan Wang, Xiaoli Feng, and Jinsong Zhang,
\newblock ``Automatic scoring at multi-granularity for l2 pronunciation.,''
\newblock in {\em Interspeech}, 2020.

\bibitem{xu2021explore}
Xiaoshuo Xu, Yueteng Kang, Songjun Cao, Binghuai Lin, et~al.,
\newblock ``Explore wav2vec 2.0 for mispronunciation detection,''
\newblock {\em Interspeech}, 2021.

\bibitem{lin2021deep}
Binghuai Lin and Liyuan Wang,
\newblock ``Deep feature transfer learning for automatic pronunciation
  assessment,''
\newblock {\em Interspeech}, 2021.

\bibitem{cincarek2009automatic}
Tobias Cincarek, Rainer Gruhn, et~al.,
\newblock ``Automatic pronunciation scoring of words and sentences independent
  from the non-native’s first language,''
\newblock {\em Computer Speech and Language}, 2009.

\bibitem{devlin2018bert}
Jacob Devlin, Ming-Wei Chang, et~al.,
\newblock ``Bert: Pre-training of deep bidirectional transformers for language
  understanding,''
\newblock {\em ACL}, 2018.

\bibitem{panayotov2015librispeech}
Vassil Panayotov, Guoguo Chen, et~al.,
\newblock ``Librispeech: an asr corpus based on public domain audio books,''
\newblock in {\em ICASSP}, 2015.

\bibitem{tu2018investigating}
Ming Tu, Anna Grabek, et~al.,
\newblock ``Investigating the role of l1 in automatic pronunciation evaluation
  of l2 speech,''
\newblock in {\em Interspeech}, 2018.

\bibitem{hernandez2018ted}
Fran{\c{c}}ois Hernandez, Vincent Nguyen, et~al.,
\newblock ``Ted-lium 3: twice as much data and corpus repartition for
  experiments on speaker adaptation,''
\newblock in {\em SPECOM}, 2018.

\bibitem{marcus1993building}
Mitchell~P Marcus, Mary~Ann Marcinkiewicz, et~al.,
\newblock ``Building a large annotated corpus of english: the penn treebank,''
\newblock {\em Computational Linguistics}, 1993.

\end{thebibliography}

\end{document}